\documentclass[sn-mathphys,Numbered]{sn-jnl}

\usepackage{graphicx}%
\usepackage{multirow}%
\usepackage{amsmath,amssymb,amsfonts}%
\usepackage{amsthm}%
\usepackage{mathrsfs}%
\usepackage[title]{appendix}%
\usepackage{xcolor}%
\usepackage{textcomp}%
\usepackage{manyfoot}%
\usepackage{booktabs}%
\usepackage{algorithm}%
\usepackage{algorithmicx}%
\usepackage{algpseudocode}%
\usepackage{listings}%
\usepackage{longtable}

\theoremstyle{thmstyleone}%
\theoremstyle{thmstyletwo}%

\theoremstyle{thmstylethree}%

\begin{document}

\title[Life Cycle Assessment of the Athena X-ray Integral Field Unit]{Life Cycle Assessment of the Athena X-ray Integral Field Unit}

\author*[1]{\fnm{Didier} \sur{Barret}}\email{dbarret@irap.omp.eu}
\author[2]{\fnm{Vincent} \sur{Albouys}}
\author[1]{\fnm{Jürgen} \sur{Knödlseder}}
\author[3]{\fnm{Xavier} \sur{Loizillon}}
\author[4]{\fnm{Matteo} \sur{D'Andrea}}
\author[5]{\fnm{Florence} \sur{Ardellier}}
\author[6]{\fnm{Simon} \sur{Bandler}}
\author[7]{\fnm{Pieter} \sur{Dieleman}}
\author[8]{\fnm{Lionel} \sur{Duband}}
\author[7]{\fnm{Luc} \sur{Dubbeldam}}
\author[4]{\fnm{Claudio} \sur{Macculi}}
\author[9]{\fnm{Eduardo} \sur{Medinaceli}}
\author[1]{\fnm{François} \sur{Pajot}}
\author[5]{\fnm{Damien} \sur{Prêle}}
\author[1]{\fnm{Laurent} \sur{Ravera}}
\author[10]{\fnm{Tanguy} \sur{Thibert}}
\author[11]{\fnm{Isabel} \sur{Vera Trallero}}
\author[1]{\fnm{Natalie} \sur{Webb}}


\affil*[1]{\orgname{Institut de Recherche en Astrophysique et Planétologie}, \orgaddress{\street{9 Avenue du Colonel Roche}, \city{Toulouse}, \postcode{31028}, \country{France}}}
\affil*[2]{\orgname{Centre National d'Etudes Spatiales}, \orgaddress{\street{18 Avenue Edouard Belin}, \city{Toulouse}, \postcode{31401}, \country{France}}}
\affil[3]{\orgname{Scalian}, \orgaddress{\street{14 Rue Paul Mesplé}, \city{Toulouse}, \postcode{31100}, \country{France}}}
\affil[4]{Istituto di Astrofisica e Planetologia Spaziali, Via Fosso del Cavaliere 100, 00133, Roma, Italy}
\affil[5]{Université Paris Cité, CNRS, CEA, Astroparticule et Cosmologie (APC), F-75013 Paris, France}
\affil[6]{NASA Goddard Space Flight Center, 8800 Greenbelt Rd, Greenbelt, MD 20771, United States}
\affil[7]{SRON, Netherlands Institute for Space Research, Niels Bohrweg 4, 2333 CA Leiden, The Netherlands}
\affil[8]{Département des Systèmes Basses Températures, CEA-Grenoble, 17 avenue des Martyrs, 38054 Grenoble cedex 99, France}
\affil[9]{INAF, Osservatorio di Astrofisica e Scienza dello Spazio, via Gobetti 93/3, 40129, Bologna, Italy}
\affil[10]{Centre Spatial de Liège, Liège Science Park, Avenue du Pré-Aily, 4031 Angleur, Belgium}
\affil[11]{INTA, Crtra de Ajalvir km 4, 28850, Torrejon de Ardoz, Madrid, Spain}



\abstract{The X-ray Integral Field Unit (X-IFU) is the high-resolution X-ray spectrometer to fly on board the Athena Space Observatory of the European Space Agency (ESA). It is being developed by an international Consortium led by France, involving twelve ESA member states, plus the United States. It is a cryogenic instrument, involving state of the art technology, such as micro-calorimeters, to be read out by low noise electronics. As the instrument was undergoing its system requirement review (in 2022), a life cycle assessment (LCA) was performed to estimate the environmental impacts associated with the development of the sub-systems that were under the responsibility of the X-IFU Consortium. The assessment included the supply, manufacturing and testing of sub systems, as well as involved logistics and manpower. We find that the most significant environmental impacts arise from  testing activities, which is related to energy consumption in clean rooms, office work, which is related to energy consumption in office buildings, and instrument manufacturing, which is related to the use of mineral and metal resources. Furthermore, business travels is another area of concern, despite the policy to reduced flying adopted by the Consortium. As the instrument is now being redesigned to fit within the new boundaries set by ESA, the LCA will be updated, with a focus on the hot spots identified in the first iteration. The new configuration, consolidated in 2023, is significantly different from the previously studied version and is marked by an increase of the perimeter of responsibility for the Consortium. This will need to be folded in the updated LCA, keeping the ambition to reduce the environmental footprint of X-IFU, while complying with its stringent requirements in terms of performance and risk management.
}




\maketitle

\section{Introduction}\label{sec1}
The Athena X-ray Integral Field Unit (X-IFU) is the high resolution X-ray spectrometer \cite{Barret_2023ExA....55..373B}, studied since 2015 for operating in the mid-30s aboard the Athena space X-ray Observatory of the European Space Agency (ESA). Athena is a versatile facility designed to address the Hot and Energetic Universe science theme \cite{2013arXiv1306.2307N,Barret_2013sf2a.conf..447B,barcons2015JPhCS.610a2008B,Barcons2017AN....338..153B,Barret2020AN....341..224B} that was selected in November 2013 by ESA's Survey Science Committee. Based on a large format array of Transition Edge Sensors (TES), X-IFU will provide spatially resolved X-ray spectroscopy, with a spectral resolution of 2.5 eV (up to 7 keV) over a hexagonal field of view of 5 arc minutes (equivalent diameter).

The X-IFU is a rather complex cryogenic instrument, involving very advanced technologies, requiring a large amount of resources and a large number of partners spread all across the world. X-IFU is being developed right at the time where 6 out of 9 planetary boundaries are already transgressed \cite{Richardson_2023SciA....9H2458R} and when it is becoming very clear that without an immediate and drastic reduction of our greenhouse gas emissions (by at least 50\% by 2030), across all sectors, the Paris Agreement goal of limiting global warming to 1.5 degrees will not be reached.
X-IFU is also developed when the need for building new large scientific infrastructures having major environmental impacts (beyond the sole greenhouse gas emission) is being debated within the community \cite{Knodlseder_2022NatAs...6..503K}, this very same community that should lead by example, because of its awareness that there is no planet B \cite{Burtscher_2022NatAs...6..764B}.

 Forging a sustainable future for astronomy requires a systemic change, and considering the environmental impact of our research projects becomes an ethical responsibility. Taking up this responsibility, the X-IFU Consortium has therefore decided to conduct a life cycle assessment (LCA) of all the components that are under its responsibility with the aim to minimise the environmental impacts related to their development.

The LCA was performed through a contract with SCALIAN, a specialized company with heritage in evaluating the environmental impact of space projects. As at the time this LCA was carried out there were very few science projects which had been subject to a LCA (see however ref.~\cite{Vargas_2023arXiv230912282V} for the LCA of the Giant Radio Array for Neutrino Detection (GRAND) experiment). On the other hand, LCA studies exist for space systems in general \cite{Wilson_2023AdSpR..72.2917W}, yet they do not provide any insights into the environmental impacts of the development of scientific payloads. Furthermore, publicly available data in the space sector are scarce and the LCAs performed so far are not yet homogenized in goal and scope definitions \cite{Wilson_2022NatAs...6..417W,Maury_2020AcAau.170..122M}. Hence it is  crucial to make LCA results publicly available that will consolidate preliminary estimates of the environmental impact of astronomical research facilities \cite{Knodlseder_2022NatAs...6..503K}. Since LCAs are new in the field of astronomical research, we included in this paper a description of the most important aspects of the LCA methodology, as defined by the ISO 14040 standard \cite{ISO14040}. 

X-IFU entered its System Requirement Review (SRR) in June 2022, at about the same time when ESA called for an overall X-IFU redesign (including the X-IFU cryostat and the cooling chain), due to an unanticipated cost overrun of Athena. While this redesign ultimately changed the configuration of the instrument, and the list of components provided by the X-IFU Consortium, we had already started our life cycle assessment on the original layout. Our assessment hence excludes components that so far were under ESA responsibility, such as the cryostat and cooling chain. For previous incarnations of X-IFU, when the cryostat and cooling chain were in the Consortium perimeter, see \cite{Barret_2013arXiv1308.6784B,Ravera_2014SPIE.9144E..2LR,Barret_2016SPIE.9905E..2FB,Barret_2018SPIE10699E..1GB,Pajot_2018JLTP..193..901P}.

\section{Method and tools}
\subsection{LCA methodology}

LCA is a standardized method that is defined by the norm ISO 14040 \cite{ISO14040} that describes the principles and framework for life cycle assessment. A LCA is conducted in three phases, comprising the definition of its goal and scope, the life cycle inventory analysis (LCI) phase, and the life cycle impact assessment (LCIA) phase. We dedicate a specific section to each of the phases (cf. sections \ref{sec:goal}, \ref {sec:lci} and \ref{sec:lcia}, respectively).
Each of the three phases is followed by an interpretation phase, with the purpose to evaluate the implications of the hypotheses taken and to check approximations, so that the reliability of the results can be improved.

The goal and scope definition phase identifies the reason for the assessment, its applications and the target audience as well as its stakeholders. It specifies in particular the system function, the functional unit, the system boundaries, as well as reference flows and key parameters.
Specifically, the functional unit is used to describe the quantified performance of a product or service, and specifies the reference unit. Defining a proper functional unit is fundamental when performing a comparative LCA, as it allows for the comparison of different scenarios or configurations that all provide the same function.

\subsubsection{Life Cycle Inventory (LCI)}
The LCI phase consists of making an inventory of the elementary input and output flows of the system under study, involving collection and organization of the required data. LCI distinguishes foreground and background data, where the former describe the system technically, while the latter describe the system environmentally. While foreground data are specific to the system under study, background data are in general extracted from LCI databases, which in our case comprise the ecoinvent database and space-system specific databases provided by ESA and SCALIAN. Matching of foreground with background data is a critical step of this phase, as relevant materials, products or processes may not be available in the background databases, requiring sometimes approximations or the development of dedicated LCI models.

\subsubsection{Life Cycle Impact Assessment (LCIA)}
\label{sec:method}
Finally, in the LCIA phase, tools are used (in our case the SimaPro software) to quantify environmental impacts according to different categories (see Table \ref{table:categories} for the categories used in this study).
This involves choosing a method for impact computation (see section \ref{sec:method}), which can broadly be classified into mid-point and end-point computation methods.
The former are used to assess the effects of emitted pollutants and resources consumed on a global scale during the life cycle, with flows of similar effects being aggregated into intermediate impact categories. For example, emissions of CO$_2$ and CH$_4$, which both contribute to global warming, can be aggregated into a climate change mid-point indicator, expressed in units of kg CO$_2$ eq.
End-point methods evaluate the potential damages that can be linked to intermediate categories, aggregating for example impacts on human health, natural resources, or biodiversity.

\label{sec:method}
We used the Environmental Footprint 3.0 (adapted) V1.01 method that is extensively used in aerospace industry and comprises 16 mid-point impact indicators, which are summarized in Table \ref{table:categories}.
The first column specifies the name of the impact indicator, the second column its unit, and the third column the detrimental effects on the environment that are captured by the category.
NMVOC stands for non-methane volatile organic compounds, CTUh for comparative toxic unit for humans, and CTUe for comparative toxic unit for ecosystems.

\begin{longtable}{|p{.20\linewidth}|p{.20\linewidth}|p{.5\linewidth}|}
\caption{Environmental impact indicators considered in the LCA of X-IFU, together with their units and their description.}
\label{table:categories} \\
\hline
\textbf{Indicator}  & \textbf{Unit} & \textbf{Description} \\
\hline 
Climate change & kg CO$_{\rm 2}$ eq & The emission of greenhouse gases into the atmosphere which absorbs and redirects heat towards earth surface. This leads to the rise of global temperature, oceans level, extreme climatic events (storm, droughts)... \\
\hline 
Ozone depletion	& kg CFC11 eq &	Some gas compounds released in the atmosphere may produce an increase of Chloride and Bromine concentration in the stratosphere. This reduces stratospheric ozone concentration for several years or decades. The effect on human health and ecosystem is premature ageing from sun UVB exposure. \\
\hline 
Ionising radiation & kBq U235 eq & The release of radioactive substances into the air and water can be inhaled or ingested by humans and ecosystems. This may lead to cancer, severe hereditary effects, long term damage on ecosystems. \\
\hline 
Photochemical ozone formation &	kg NMVOC eq & Creation of ozone in the troposphere (where it is not supposed to) due to the reactions between sun UV rays and VOC/nitrogen oxides emitted. This can lead to respiratory distress for mammals and reduction of photosynthesis for ground and water based plants. \\
\hline 
Particulate matter & disease incidences & The emission of particular matter or their precursors (nitrous oxides, Sulphur oxides and ammonia) can trigger respiratory diseases. \\
\hline 
Human toxicity, non-cancer & CTUh & Emission of chemicals into the environment, combined with the exposition of humans, taking into account absorbed dosis and risk potential. This leads to damage of human health. \\
\hline 
Human toxicity, cancer & CTUh & Emission of chemicals into the environment, combined with the exposition of humans, taking into account absorbed dosis and risk potential. This leads to damage of human health. \\
\hline 
Acidification & mol H$^+$ eq & Emission into the atmosphere of substances with acidic effects (nitrogen oxides, Sulphur oxides, and ammonia) and deposition of these substances on the ground, that increase the acidity of soils and water. This leads to a decrease of biodiversity and bio-productivity. \\
\hline 
Eutrophication, freshwater & kg P eq & Release of nitrous and phosphorous substances into the ecosystems. This increases the amount of nutrients available, which leads the proliferation of microorganisms that reduce the oxygen concentration and thus biodiversity. \\
\hline 
Eutrophication, marine & kg N eq & Release of nitrous and phosphorous substances into the ecosystems. This increases the amount of nutrients available, which leads the proliferation of phytoplankton that reduce the oxygen concentration and thus biodiversity. \\
\hline 
Eutrophication, terrestrial	& mol N eq  & Release of nitrous and phosphorous substances into the ecosystems. This leads to modification of plants’ nutritive balance, which decreases their productivity. \\
\hline
Ecotoxicity, freshwater & CTUe & Emission of chemicals into the environment combined to the exposition of freshwater species, taking into account absorbed dosis and vulnerability of each species. This leads to damage to ecosystems. \\
\hline 
Land use & Pt & Conversion of land for agriculture, resource extraction or dwelling. This leads to physical changes upon the environment that can modify species distributions. \\
\hline 
Water use & m$^3$ depriv. & The use of water in a specific quantity and quality may trigger usage competition: agriculture, domestic consumption, fishing and ecosystems. \\
\hline 
Resource use, fossils & MJ & The use of primary energy is mainly responsible for fossil resource depletion. \\
\hline 
Resource use, minerals and metal & kg Sb eq & The extraction of mineral and metal resources reduce their availability for further use. \\
\hline 
\end{longtable}

\subsection{Normalisation, weighting and single score}
LCIA results may be presented in absolute units (as specified in Table \ref{table:categories}), normalised to a common reference, or weighted so that different impact categories can be aggregated into a single impact score.
Normalising and weighting depends on the LCIA method chosen, which in our case was the Environmental Footprint 3.0 (EF3.0) method (see section \ref{sec:method}).
The normalisation factors used in this method are
the environmental impacts caused by an average human being, without considering geographical region, based on the world population in 2010 \cite{sala2017}.

Weighting consists of attributing to each impact category a unit rating that represents the global environmental footprint of the studied system. In that way, different impact categories can be aggregated into a single impact score.
Weightings are computed against several criteria, such as current state of science or LCA practitioner.
As such, weightings introduce a significant and subjective uncertainty, and consequently, results should be interpreted with care. For example, EF3.0
considers climate change as particularly important, and hence associates a large weight to this category, while effects on toxicity and biodiversity have a smaller weight. EF3.0 normalisation and weighting factors that were adopted in this study are summarized in Table \ref{tab:factors}.

\begin{table}[!h]
\centering
\begin{tabular}{l|c|c}
{\bf Indicator} & {\bf Normalisation factor} & {\bf Weighting factor} \\
\hline
Climate change & 0.0001235 & 0.2106 \\
Ozone depletion & 18.64 & 0.0631 \\
Ionising radiation & 0.0002370 & 0.0501 \\
Photochemical ozone formation & 0.02463 & 0.0478 \\
Particulate matter & 1680 & 0.0896 \\
Human toxicity, non-cancer & 4354 &	0.0184 \\
Human toxicity, cancer & 59173 & 0.0213 \\
Acidification & 0.01800 & 0.062 \\
Eutrophication, freshwater & 0.6223 & 0.028 \\
Eutrophication, marine & 0.05116 & 0.0296 \\
Eutrophication, terrestrial & 0.005658 & 0.0371 \\
Ecotoxicity, freshwater & 0.00002343 & 0.0192 \\
Land use & 0.000001220 & 0.0794 \\
Water use &	0.00008719 & 0.0851 \\
Resource use, fossils & 0.00001538 & 0.0832 \\
Resource use, minerals and metal & 15.71 & 0.0755 \\
\hline
\end{tabular}
\caption{Normalisation and weighting factors of the EF3.0 method that were adopted in this study.}
\label{tab:factors}
\end{table}

\subsection{Tools}
The LCA was conducted using the SimaPro software version 9.3.0.3, and the life cycle inventory was matched with specific space-systems databases from ESA (version 3) and SCALIAN, and the ecoinvent database version 3.7.1. The ESA and the SCALIAN databases collect LCA information that was used as background information for aviation and space-sector specific processes (including for example space qualified PCBs and metals for the former, and test facilities and logistics for the latter). The ecoinvent database was used for standard industrial processes and materials, and served as background database for the space-sector specific databases.

\subsection{Data collection and quality}
\label{sec:quality}
X-IFU is still in an early development phase, implying that the project is still subject to changes in schedule and responsibility, which may lead to revision of scope or assumptions in the future. 
Even though main technologies have been defined, the supply chains are not always completely identified, preventing the collection of accurate data on manufacturing processes and intermediate transportation flows.
Due to the unavailability of such primary data, we relied in this study on secondary or generic data that were collected through dedicated interviews with sub-system leaders.

To consider the uncertainties related to the data collection, we defined a confidence scale to trace the quality of the foreground data, and its correspondence to entries in the background database (see Table \ref{tab:quality}).

\begin{table}[]
\centering
\begin{tabular}{|l|c|}
\hline
{\bf Foreground data accuracy} & {\bf Score} \\
\hline
No data provided & 0  \\
Unreliable data, rough estimate, remote proxy & 1  \\
Proxy estimate & 2  \\
Reliable data based on expert assumption or literature & 3  \\
Direct measurement & 4  \\
\hline
{\bf Correspondence with background database entry} & {\bf Score} \\
\hline
No entry found that can be remotely linked to the flow data & 0  \\
The entry is a proxy that could be representative (estimate) & 1  \\
The entry is a proxy from the same process/material family & 2  \\
Close or similar process/material & 3  \\
Perfect match & 4  \\
\hline
\end{tabular}
\caption{Confidence scale used to classify the quality of the foreground data and their correspondence with entries of the background database.}
\label{tab:quality}
\end{table}



\section{Goal and Scope definition}
\label{sec:goal}

\subsection{Goal of the assessment}

The aim of the LCA is the identification of significant environmental impacts arising from the activities of the X-IFU Consortium, with the goal to take corrective actions for impact reductions while not jeopardizing the ultimate success of the project.
The X-IFU Consortium had already committed to reduce significantly its global travel footprint \cite{Barret_2020ExA....49..183B}, starting even in the pre-covid era, but this LCA should go one step further, by identifying additional actions that can lead to substantial reductions in the environmental impacts of X-IFU Consortium activities.

\subsection{Functional unit}

The functional unit defined for this study is “Provision of all X-IFU components to ESA according to requirements, except of the cryostat and cooling chain, enabling spatially resolved high-resolution X-ray spectroscopy aboard the Athena satellite during a nominal lifetime of 4 years, yet with design rules enabling the mission to operate 10 years in orbit”.

\subsection{Boundaries of the system}
\begin{figure}[!h]
\centering
\includegraphics[width=13cm]{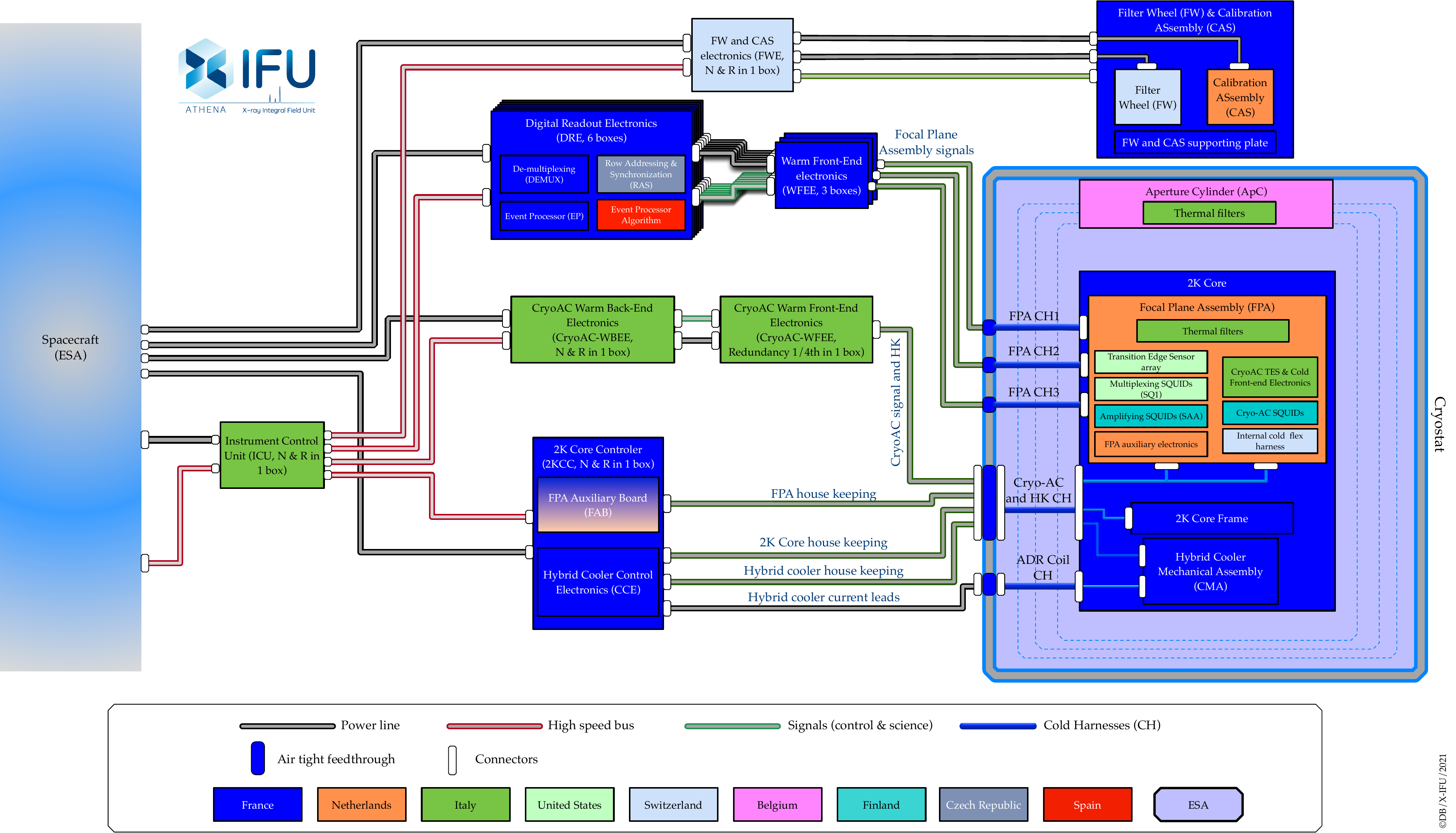}
\caption{The physical breakdown of the X-IFU instrument, highlighting its main components, as well as the country responsible for the procurement \cite{Barret_2023ExA....55..373B}. Elements shown as in the perimeter of ESA were excluded from the LCA of X-IFU. This breakdown was the baseline for the system requirement review held in 2022.}
\label{fig:phy_breakdown}
\end{figure}

This LCA was restricted to components placed under the responsibility of the X-IFU Consortium (in the version of the X-IFU instrument extensively described in \cite{Barret_2023ExA....55..373B}), and illustrated in Figure \ref{fig:phy_breakdown}, which represents a physical and programmatic breakdown of the instrument.
Accordingly, the focal plane assembly and its internal components, such as the transition edge sensors, the sub-kelvin cooler, the control and readout electronics, as well as ancillary equipments, such as filter wheel and calibration assembly, are included in the assessment.
Other elements necessary to implement the X-IFU instrument, as well as the Athena mission in general, such as the cryostat, the cooling chain (with the exception of the sub-kelvin cooler), the payload module, the X-ray telescope and other instruments, the spacecraft and the launch vehicle, and related integration activities, were not considered in this study.
Table \ref{table:systems} summarises the sub-systems that were considered in this study together with the geographical locations of manufacturing that were used in the LCI modelling.

The LCA is hence limited to X-IFU Consortium activities, over a period spanning the years 2015 to 2034, covering mission phases B-D, excluding later phases that cover launch preparation, launch, operations, science activities, and end-of-life processes.
Specifically, the LCA determines the environmental impacts of X-IFU Consortium deliverables from cradle to gate, where the gate corresponds to the delivery of the X-IFU components to ESA for integration.
The assessment included the supply, manufacturing and testing of sub-systems, as well as involved logistics and manpower.

\begin{table}[]
\centering
\begin{tabular}{|l|l|c|}
\hline
{\bf Subsystem} & {\bf Description} & {\bf Country} \\
\hline
2KCC HybCCE & Hybrid Cooler Electronics & France \\
Amp SQUID & Amplification SQUIDs & Finland \\
ApC & Aperture Cylinder & Belgium \\
CAS & Calibration Assembly & Netherlands \\
2K Core & 2 Kelvin core & France \\ 
CryoAC & Cryogenic Anticoincidence Detector & Italy \\
Test bench & Cryogenic test bench & France \\
DRE & Digital Readout Electronics & France \\
EBIT & Electron Beam Ion Trap & France \\
EP Algorithm & Event Processor Algorithm & Spain \\
FW & Filter Wheel & Switzerland \\
FPA & Focal Plane Assembly & Netherlands \\
ICU & Instrument Control Unit & Italy \\
SQUIDs & Multiplexing SQUIDs & USA \\
Simulator & Scientific simulator & Germany \\
SubK Cooler & Sub-Kelvin cooler & France \\
TES Array & Transition Edge Sensor Array & USA \\
TGSE & Thermal Ground Support Equipment & Spain \\
THF & Thermal filters & Italy \\
WFEE & Warm Front-End Electronics & France \\
\hline
\end{tabular}
\caption{X-IFU sub-systems considered in this study.}
\label{table:systems}
\end{table}

\section{Life Cycle Inventory}
\label{sec:lci}

\subsection{Data collection}
Foreground data of the LCI were collected through one-to-one interviews with subsystem project leaders or their representatives. 
The aim of these interviews was to get a comprehensive view on the materials and processes, as well as the necessary conditions, for subsystems manufacturing and testing.
While data collection was iterated for all subsystems, not all subsystems were at the same definition or maturity level, leading to heterogeneous uncertainties in the collected foreground data.
The collected foreground data were classified into the following four categories:

\begin{itemize}
\item {\bf Manufacturing}, comprising the manufacturing aspects of the instrument, such as raw materials, parts, Commercial Off-The-Shelf (COTS) items, manufacturing processes, number of intermediate models and packaging.
\item {\bf Logistics}, covering all steps necessary to ensure subsystems transportation to testing locations and the X-IFU integration site at Centre National d'Etudes Spatiales (CNES) premises.
\item {\bf Testing}, including information on the use of testing equipment that can involve energy or matter, such as the use of clean rooms, vacuum chamber, shakers or cryogenic coolers.
\item {\bf Manpower}, covering the number of people involved in the activities and their travelling due to X-IFU Consortium.
\end{itemize}
In the following, the data collection for these four categories are explained in some detail.
Uncertainties in the data collection, as discussed below, were handled through the confidence scale described in section \ref{sec:quality}.

\subsection{Manufacturing}
The accuracy of manufacturing data that were available depend on the maturity level and development stage of the subsystem, eventually due to design evolution related to project constraints, or due to an undefined manufacturing process that may depend on the yet to be decided industrial supplier.

As illustration for material used, information may vary between ``Aluminum 6062" (very detailed), ``Aluminum 6XXX family" (detailed), ``Aluminum box" (rough), and ``Metal" (uncertain). Some materials were also described as COTS so only partial public information could be recovered.

The accuracy with which the manufacturing processes are known was also variable. For example, information on surface treatment may vary between ``heat treatment followed by black anodizing" (accurate), ``surface treatment" (approximate), and ``no information".
 
The buy-to-fly ratio, which is the amount of material bought versus the amount of material remaining on the final product, is also very variable, and depends very much on the maturity of the system to build. For example, buy-to-fly ratios may vary between ``50 kg bought for 2 kg remaining after milling" (accurate), ``around twice the mass" (rough), and ``no information".
Most of the buy-to-fly ratios were actually unknown, and considered as being equal to one when no information was provided.

Data on packaging were rarely available as they are highly dependent on the supplier. 
Based on feedback from other LCAs of aerospace components, as packaging impact is often low ($<5\%$ of overall LCA single score), it was not included nor extrapolated in this first iteration of the LCA.
There is significant room for improvement in the next LCA iteration as resource use contributes significantly to the environmental impacts of space projects.

\subsection{Logistics}
The necessary transportation steps to move subsystems from one location to another or to the final integration site have been estimated even though they may evolve with time. Mostly air and road transportation were identified at this stage.

\subsection{Testing}
Testing activities were described following the equipment inventory carried out during the interviews. For most subsystems a clean room has to be used for the whole testing duration. The test campaign can be several months long even if the test equipment might only be used actively for a few hours. Data collected to describe facilities and equipment involved are highly variable, as in most cases, the facilities are managed by an entity external to the team. Four items were considered for the testing:
\begin{itemize}
\item Equipment type;
\item Nominal power for active and stand by phases, and the duration of these phases;
\item Consumption (helium, water);
\item Test duration.
\end{itemize}
When data provided were not accurate enough, assumptions were made based on information of other subsystems or feedback from other space projects carried out at CNES.

The impact from processing and storage of test data should in principle also be included in this category, but no extensive data use dedicated to X-IFU was identified during the interviews.
Data use was therefore neglected in the analysis.

\subsection{Manpower}
Manpower contributes to the environmental impact of a project due to emissions from office buildings hosting the people (electricity use, heating, air conditioning, waste management, water consumption), emissions related to purchase of basic goods and services (furniture, computers, screens, office supplies, internet, telecommunications), business travels, home-to-office commuting and lunches. 
While most of these emissions are shared by different activities and projects, we follow the ESA LCA guidelines and estimate the fraction attributable to X-IFU by the ratio of Full Time Equivalents (FTE) dedicated to the project compared to the total FTE working in the premises \cite{ESAHandbook}.

We base our estimate on the inventory of fluxes occurring during the daily operations of the Institut de Recherche en Astrophysique et Planétologie (IRAP) in the year 2019 that were collected in the context of the carbon footprint estimate of the laboratory \cite{martin2022}.
In order to extract the environmental impacts that are generated by hosting staff personal in the office buildings, we exclude all fluxes that relate to project and science activities, avoiding hence possible double-counting of environmental impacts in the manpower category.
Specifically, we consider electricity consumption, heating of office buildings, leakage of refrigerants from air conditioning, water consumption and waste water treatment, waste treatment, purchase of laptops, and home-to-office commuting, excluding other fluxes such as purchase of goods and services, business travelling, and use of observatory data.
In total, 263 FTE worked at IRAP in 2019, which results in a carbon intensity of 2.7 tCO${_2}$eq/FTE for the selected fluxes, which is about 10\% of the 28 tCO${_2}$eq/FTE that is obtained when also project and science related fluxes are included.

We specify in Table \ref{sec:ftemodel} the inventory of fluxes that we adopted for 263 FTE working at IRAP.
Based on this inventory, we created country-dependent LCI models for manpower by selecting the relevant country-specific processes from the ecoinvent database, in particular the country-specific electricity mix and waste treatment impacts.
This leads to carbon intensities, ordered by increasing value, of
2.2 tCO${_2}$eq/FTE for Switzerland,
2.7 tCO${_2}$eq/FTE for France,
3.7 tCO${_2}$eq/FTE for Belgium,
4.0 tCO${_2}$eq/FTE for Finland,
4.4 tCO${_2}$eq/FTE for Spain,
5.3 tCO${_2}$eq/FTE for Italy,
5.8 tCO${_2}$eq/FTE for Germany,
6.2 tCO${_2}$eq/FTE for the Netherlands and the United States, and
10.5 tCO${_2}$eq/FTE for Poland.
To assess the manpower impacts, information was gathered from sub-system project leaders on the size of the team dedicated to X-IFU and how they foresaw its evolution over the project duration.

\begin{table}[!t]
\centering
\begin{tabular}{l|c|c}
{\bf Process} & {\bf Unit} & {\bf Quantity} \\
\hline
Electricity consumption & kWh & 2\,276\,000 \\
Heating (natural gas) & MJ & 774\,000 \\
Heating (wood) & MJ & 3\,085\,200 \\
Air conditioning (HFC-152a leakage) & kg & 14.06 \\
Water consumption & kg & 4\,744\,000 \\
Waste water treatment & m$^3$ & 4744 \\
Waste treatment & kg & 155\,000 \\
Purchase of laptops & unit & 139 \\
Home-to-office commuting (passenger train) & pkm & 254\,503 \\
Home-to-office commuting (regular bus) & pkm & 120\,105 \\
Home-to-office commuting (tram) & pkm & 74\,583 \\
Home-to-office commuting (small size petrol car) & km & 299\,985 \\
Home-to-office commuting (small size diesel car) & km & 474\,806 \\
Home-to-office commuting (small size natural gas car) & km & 6\,396 \\
Home-to-office commuting (electric car) & km & 96\,304 \\
Home-to-office commuting (electric bicycle) & km & 33\,136 \\
Home-to-office commuting (motor scooter) & km & 32\,733 \\
\hline
\end{tabular}
\caption{Inventory of fluxes for 263 FTE working at IRAP in 2019.}
\label{sec:ftemodel}
\end{table}

In addition to the impacts of office work, we also assessed the impacts of projected related business travelling.
Business travels by train or by plane related to X-IFU activities have been assessed from budget forecasts for each sub-system.
A pre-pandemic scenario has been considered, which, given the observed post-pandemic trends, probably implies that the impact of business travels will be overestimated.

\section{Life Cycle Impact Assessment}
\label{sec:lcia}

\subsection{Absolute impacts}
The absolute environmental impacts that we computed from the LCI are summarised in Table \ref{sec:absolute}.
We estimate a total carbon footprint of 25.5 ktCO$_2$eq associated with the development, testing and delivery of X-IFU sub-systems.
The mass of the sub-systems that were considered in this assessment amounts to 221.3 kg \cite{Barret_2023ExA....55..373B}, resulting in an emission factor of 115.2 tCO$_2$eq per kg of instrument mass, about twice the value of 50 tCO$_2$eq per kg of satellite wet mass that was inferred by ref.~\cite{Knodlseder_2022NatAs...6..503K} from LCAs of space missions.
We note, however, that the latter value refers to an entire space mission, combining elements that were not assessed in our study (full payload, satellite, launcher, ground operations).
Furthermore, the mass reference differs (instrument mass versus satellite wet mass), hence the values are strictly speaking not comparable.
Still, the relatively large emission factor may indicate that scientific instrumentation eventually contributes significantly to the footprint of a scientific space mission, strengthening the necessity for a detailed evaluation of its environmental impacts and vigorous actions for their reductions.

\begin{table}[!t]
\centering
\begin{tabular}{l|c|c}
{\bf Indicator} & {\bf Unit} & {\bf Total} \\
\hline
Climate change & kg CO$_2$ eq & 25\,536\,128 \\
Ozone depletion & kg CFC11 eq & 2.299 \\
Ionising radiation & kBq U235 eq & 15\,945\,955 \\
Photochemical ozone formation & kg NMVOC eq & 81\,952 \\
Particulate matter & desease incidents & 1.001 \\
Human toxicity, non-cancer & CTUh & 0.454 \\
Human toxicity, cancer & CTUh & 0.01164 \\
Acidification & mol H$^+$ eq & 109\,898 \\
Eutrophication, freshwater & kg P eq & 11\,615 \\
Eutrophication, marine & kg N eq & 27\,103 \\
Eutrophication, terrestrial & mol N eq & 267\,762 \\
Ecotoxicity, freshwater & CTUe & 401\,284\,747 \\
Land use & Pt & 200\,962\,056 \\
Water use &	m$^3$ depriv. & 5\,910\,066 \\
Resource use, fossils & MJ & 598\,393\,056 \\
Resource use, minerals and metal & kg Sb eq & 746.4 \\
\hline
\end{tabular}
\caption{Absolute LCIA results.}
\label{sec:absolute}
\end{table}

\begin{figure}[!t]
\centering
\includegraphics[width=13cm]{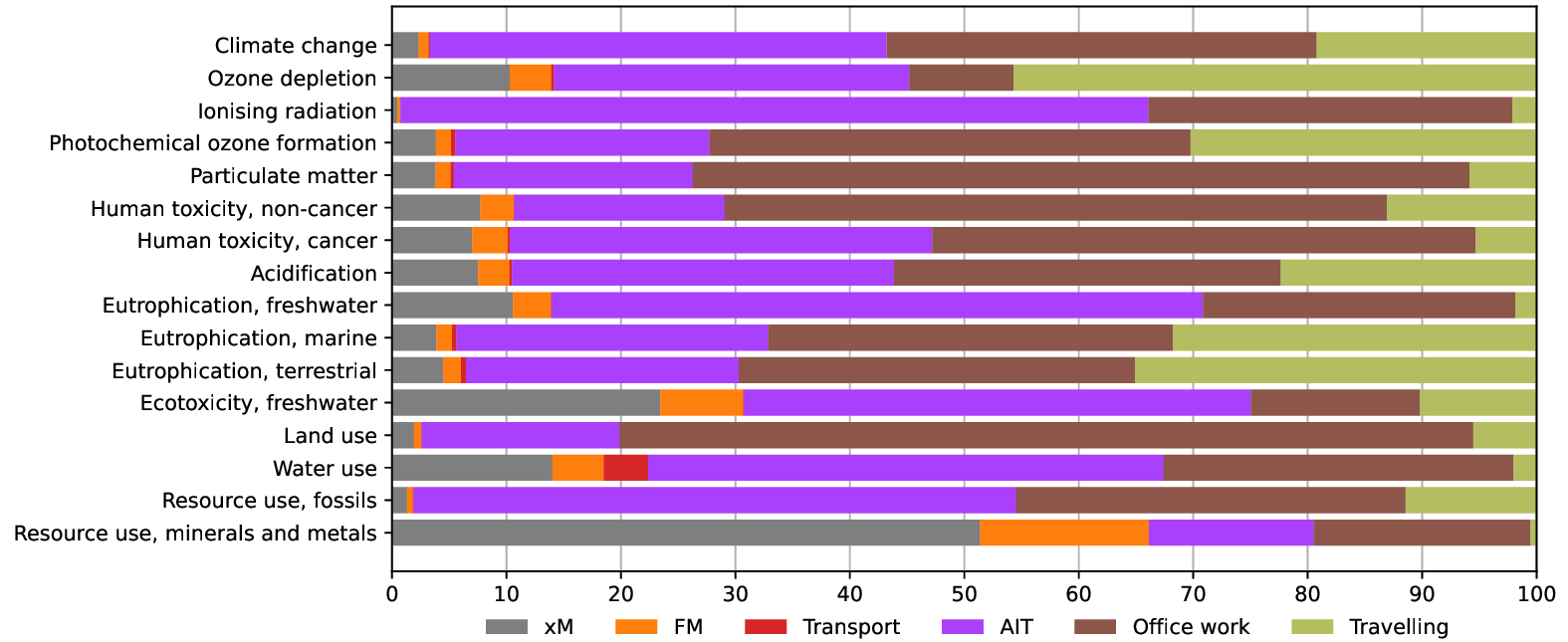}
\caption{Contribution of activities to the environmental impacts of X-IFU.}
\label{fig:contributions}
\end{figure}

The relative contributions of the different products and activities to the environmental impact of X-IFU are illustrated in Figure \ref{fig:contributions}.
For the latter, impacts have been split into qualification models (xM), such as manufacturing of structural thermal models, engineering models, qualification models, proto-flight models, and spares, the Flight Model (FM), Transport, Assembly, Integration and Testing (AIT), Office work, and Travelling.

Clearly, for most impact categories, AIT, office work and business travels contribute most significantly to the environmental impacts of X-IFU.
AIT has particular large contributions to ionising radiation, owing to the use of nuclear energy for electricity production in France, as well as to freshwater eutrophication, freshwater ecotoxicity, water use and use of fossil resources, all related to operating clean rooms.
The impact category that stands out with substantially different contributions is the use of mineral and metal resources, which is dominated by the manufacturing of qualification or flight models.

Office work contributes also significantly to many impact categories, a result that confirms the findings of ref.~\cite{chanoine2017} who assessed the environmental impacts of the Sentinel 3B mission.
We observe particular important contributions to particulate matter, non-cancer and cancer human toxicity, and land use.
Particulate matter and land use are related to heating of office buildings by wood, which is the case for the IRAP building at the Roche-site, and hence an artifact of the specific LCI model that we used to estimate the impacts of office work.
Human toxicity impacts arise from electricity consumption in the office buildings.

Business travelling also contributes substantially to the environmental impacts, in particular for ozone depletion, photochemical ozone formation, and marine and terrestrial eutrophication.
On the other hand, transport of sub-systems contributes little to the environmental impacts of X-IFU, with the exception of a $\sim3\%$ contribution to water use.
Also the manufacturing of the flight model contributes only modestly to the environmental impacts, with the exception of the use of mineral and metal resources that is dominated by instrument manufacturing.

\subsection{Normalised life cycle impacts}
\begin{figure}[!h]
\centering
\includegraphics[width=13cm]{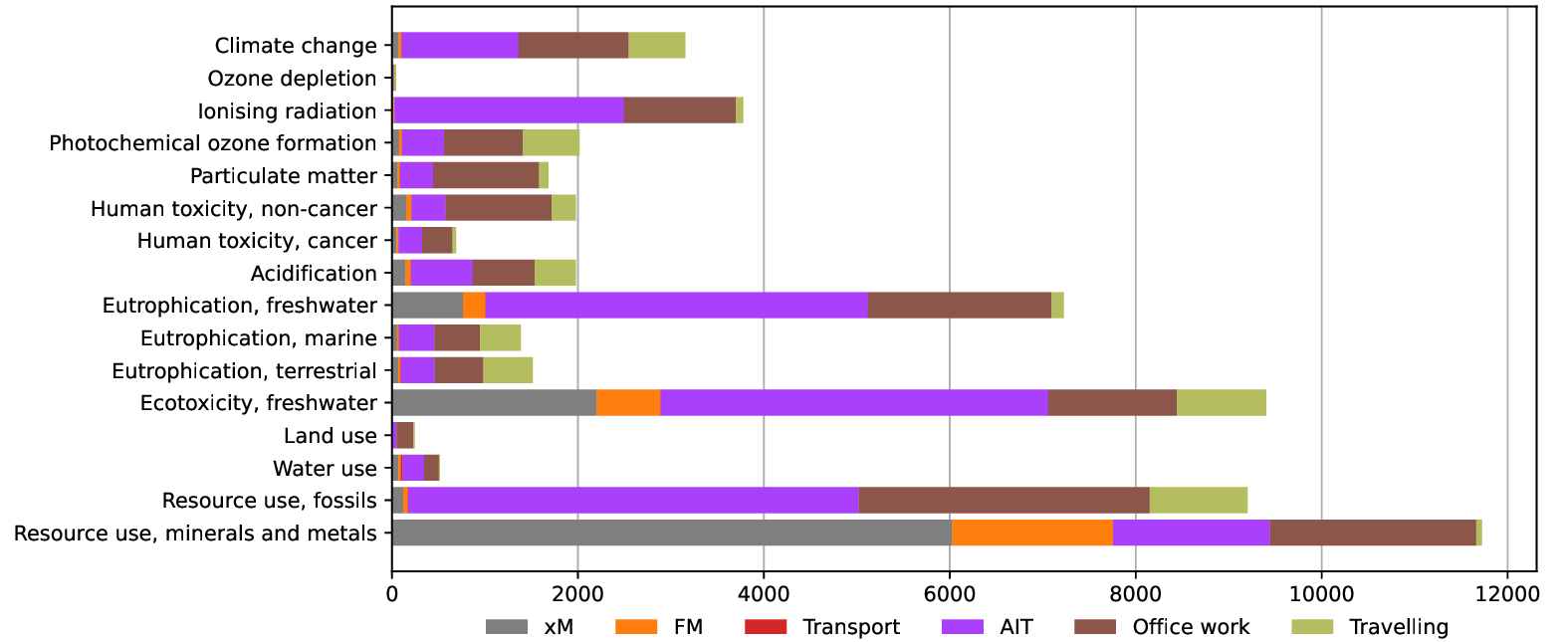}
\caption{Normalised environmental impacts of X-IFU.}
\label{fig:normalised}
\end{figure}

To compare the different impact categories, we show in Figure \ref{fig:normalised} the normalised impacts, using the normalisation factors given in Table \ref{tab:factors} for the EF3.0 method.
By multiplying absolute impacts with the normalisation factors, environmental impacts are expressed in units of annual impacts of an average human. 
For example, a normalised value of 1000 corresponds to the annual impact generated by 1000 average humans in the world, where average refers to the ignorance of impact inequalities among individual humans.

The most significant impact category is mineral and metal resources use, with impacts being primarily related to the manufacturing of instrument models.
The next most significant impact category is freshwater ecotoxicity, with impacts arising both from the manufacturing of instrument models as well as from AIT, office work and business travelling.
Next comes use of fossil resources, which is largely dominated by AIT, office work and business travelling, and freshwater eutrophication, again with a mix of instrument model manufacturing, AIT, and office work.

\subsection{Impact weighting}
\begin{figure}
\centering
\includegraphics[width=13cm]{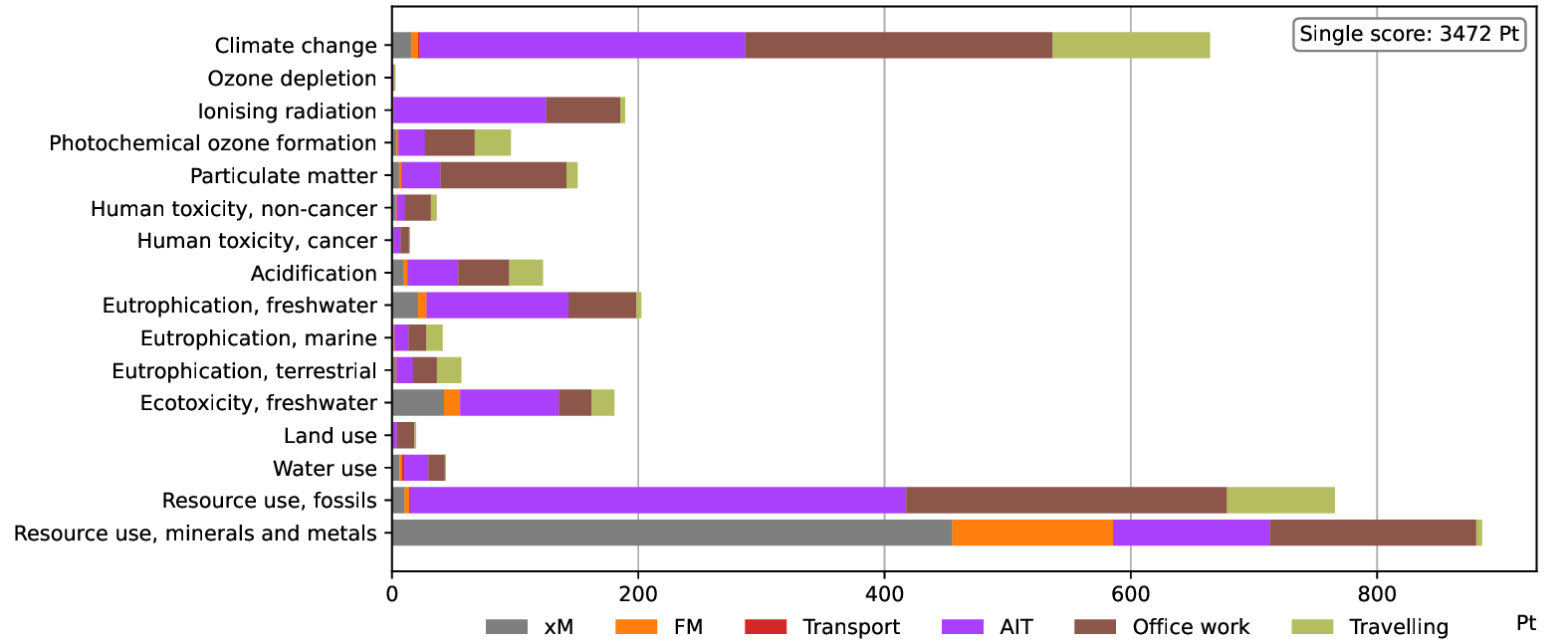}
\caption{Weighted environmental impacts of X-IFU.}
\label{fig:weighting}
\end{figure}

As alternative representation, we show in Figure \ref{fig:weighting} the weighted impacts as derived using the normalisation and weighting factors given in Table \ref{tab:factors} for the EF3.0 method.
This analysis suggests that climate change and resource use, including fossil fuels as well as mineral and metal resources, are the most significant environmental impacts of X-IFU, followed by freshwater eutrophication, ionising radiation, and freshwater ecotoxicity.
We recall that weighted impacts depend on expert judgments on the importance of each impact category, introducing some subjectivity in the assessment.
Yet through weighting, environmental impacts are put on a common scale (called ``points''), allowing for aggregation of results among impact categories.
Adding up the weighted impacts, we obtained a single score of 3472 points, which means that the environmental impacts of the X-IFU systems considered in this study correspond to the annual environmental load of 3472 humans.

\begin{figure}
\centering
\includegraphics[width=13cm]{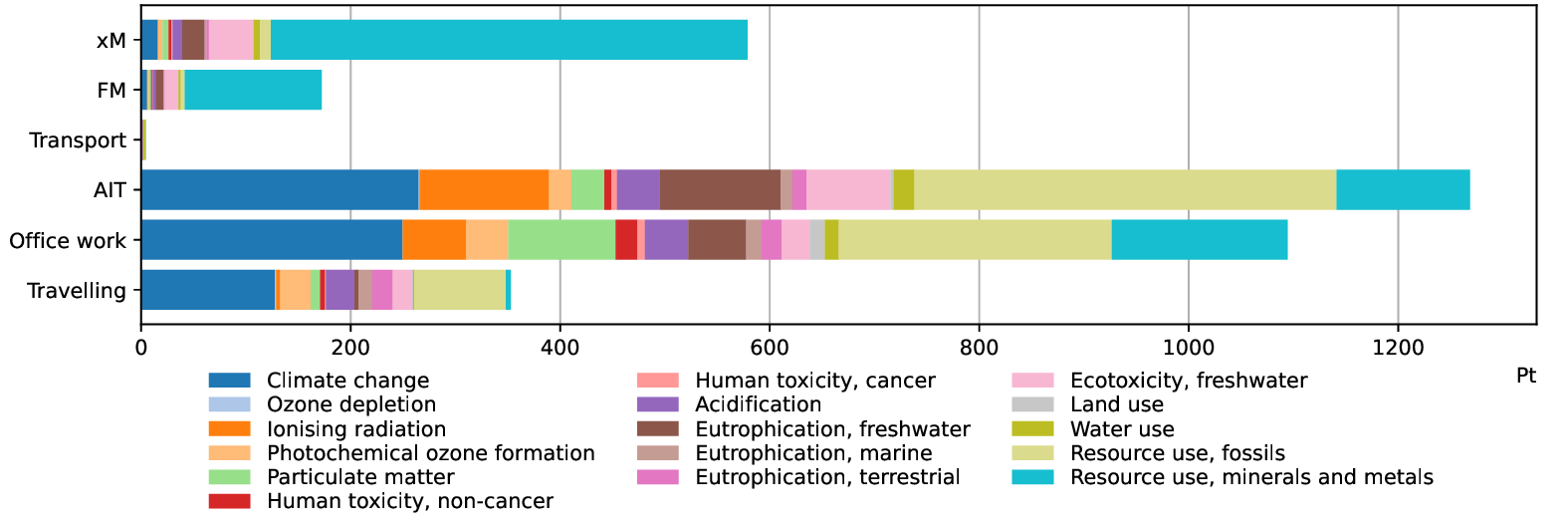}
\caption{Single score environmental impacts of X-IFU.}
\label{fig:singlescore}
\end{figure}

In Figure \ref{fig:singlescore} we break down the aggregated impacts by category, which suggests that AIT and office work contribute most significantly to the environmental impacts of X-IFU.
Both aggregate significant environmental impacts from many different categories, with fossil resource use and climate change contributing to about half of the impacts.
The manufacturing of instrument model is the third most impacting activity, which is clearly dominated by use of mineral and metal resources.
Next comes travelling, with impacts in many categories.
Finally, environmental impacts from transport appear to be negligible.

\begin{figure}
\centering
\includegraphics[width=13cm]{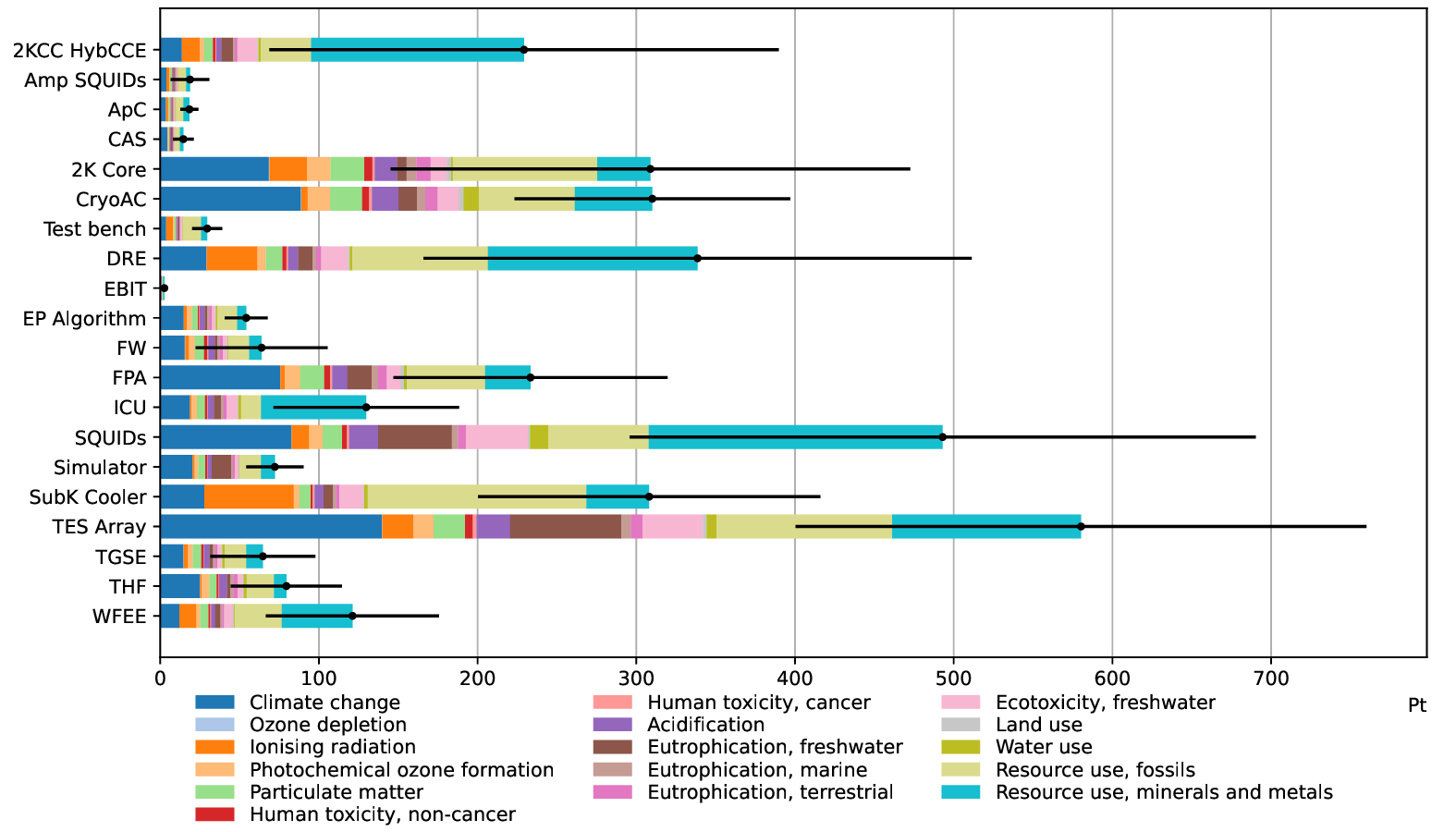}
\caption{Single score environmental impacts of X-IFU sub-systems. The horizontal bars indicate the uncertainty in the total environmental impact per subsystem.}
\label{fig:subsystems}
\end{figure}

Figure \ref{fig:subsystems} shows the aggregated environmental impacts by subsystem (cf.~Table \ref{table:systems}).
Uncertainties in the total environmental impacts per sub-system are indicated by black bars.
Uncertainties range from 25\% to 70\%, depending on the quality of data collection and the maturity level of the sub-system.
The mean uncertainty in the estimates is about 40\%, which is acceptable, given the maturity of the project.
Prime contributors to the environmental impacts are the TES Array and the multiplexing SQUIDs.
Next come the Digitial Readout Electronics, the 2 Kelvin core, the Cryogenic Anticoincidence Detector and the sub-Kelvin cooler which all have comparable impacts around 300 points.

\section{Discussion} 
The manufacturing of X-IFU hardware, including pre-flight (xM) and flight models (FM), contribute about 22\% to the environmental impacts of X-IFU. 
Transport of subsystems turns out to have a negligible environmental impact, below 1\%, probably related to the relatively small weight of the equipment. 
Testing contributes to about 37\% to the environmental impacts of X-IFU, which is explained by the significant amount of energy used over time for testing related activities, in particular in relation to the use of clean rooms. 
Office work contributes to about 32\% to the environmental impacts of X-IFU, owing primarily to power and heating office buildings and commuting of personnel to the laboratory. 
Business traveling by plane is another category with significant contribution to X-IFU’s environmental impact, estimated to be about 10\%. Next, we provide some more details about each of these items.

\subsection{Manufacturing}

The manufacturing of X-IFU hardware, including pre-flight (xM) and flight models (FM), contribute significantly to the environmental impacts of X-IFU.
X-IFU relies on high performance state of the art technologies and electronics, incorporating precious metals and rare earth elements  which are known for their very low availability and polluting refining processes.
The latter contribute to environmental impacts in many categories, including in particular freshwater ecotoxicity and eutrophication.
It is worth stressing that manufacturing impacts are very likely underestimated, as most buy-to-fly ratios, which can be very high for space applications, are not available for most X-IFU subsystems.
Getting more accurate inputs will be the challenge of the next updates of the X-IFU LCA, specifically when major hardware suppliers are selected.
The model philosophy applying to X-IFU is intended to mitigate the risks associated with the development of the instrument, and it is very unlikely that intermediate models can be retired.
Nevertheless, the Protoflight Model (PFM) approach followed by the project means that PFM models of sub-systems (instead of Engineering Qualification Models) will become Flight Spares (fully functional with qualified parts, processes and materials), thus saving one model production.
In all cases, it appears that the impact of the model philosophy is tightly connected to the risk management approach, which for expensive space projects is highly constrained and often the key point in the decision process form committing the large spending associated with the project.
This means that no major reduction of the environmental impacts associated with the model manufacturing can be anticipated in projects like X-IFU, yet questioning and challenging the number of intermediate units produced should be considered, whenever deemed acceptable.

\subsection{Logistics}
Transport of sub-systems turns out to have a negligible environmental impact, probably related to the relatively small weight of the equipment.
The small contribution of transport to the overall impacts of X-IFU should be verified once more detailed data on suppliers is available and the integration logic of the instrument is fully defined.

\subsection{Testing}
A significant amount of energy is used over time for testing related activities, in particular in relation to the use of clean rooms.
Depending on the electricity mix of the country where the task is carried out, the most significant resulting impacts will be climate change, eutrophication, ecotoxicity and fossil resource use for fossil-based electricity mixes, while ionizing radiation will dominate for nuclear electricity production, which dominates in France.
As X-IFU is still in its early development phase there is room for optimizing the testing strategy, although acceptance and qualification tests are necessary for managing the risks of the project.
Yet the duration of the tests for each model, as well as duplication of tests at different levels of instrument integration could eventually be streamlined, reducing the related environmental impacts.
The energy usage of the clean rooms bears also potential for optimisation, as well as the usage of disposable consumables needed for clean room operations.
Eventually, test campaigns could be relocated to facilities that have a demonstrable lower environmental impact, reducing thus the overall footprint of the project.

\subsection{Office work}
Office work contributes significantly to the environmental impacts of X-IFU, owing primarily due to power and heating office buildings and commuting of personal to the laboratory.
Lever arms to reduce the environmental impacts of office work include energy savings in the office buildings, eventually linked to thermal insulation of the building in case that heating and cooling are significant contributors to energy consumption.
Furthermore, reduction of purchases of new material, either by extending the lifetime of equipment or by purchasing refurbished or recycled goods, will reduce the environmental footprint.
Home-to-office commuting, as well as midday meals are further areas where environmental impact reductions can be achieved, for example by promoting active mobility (walking, cycling) or public transport instead of use of individual cars, and by offering less red meat (beef, sheep) as well as vegetarian and vegan options in the office canteens.

\subsection{Travelling}
Business travelling by plane is another category with significant contribution to X-IFU's environmental impact.
We recall that a pre-pandemic scenario has been considered in this study, and given the observed post-pandemic trends, the contribution of this category may be actually lower.
In particular, the X-IFU Consortium has already committed to reduce significantly its global travel footprint, for example by reducing the number of large consortium meetings to one per two years, and progressively replacing face-to-face working meetings by video conferences  \cite{Barret_2020ExA....49..183B}.
It remains to be seen how these commitments impact the overall project footprint, and whether additional actions can be taken to reduce X-IFU's travelling footprint.

\section{Revisiting the LCA of the new X-IFU}
\begin{figure}[!t]
    \centering
    \includegraphics[width=13cm]{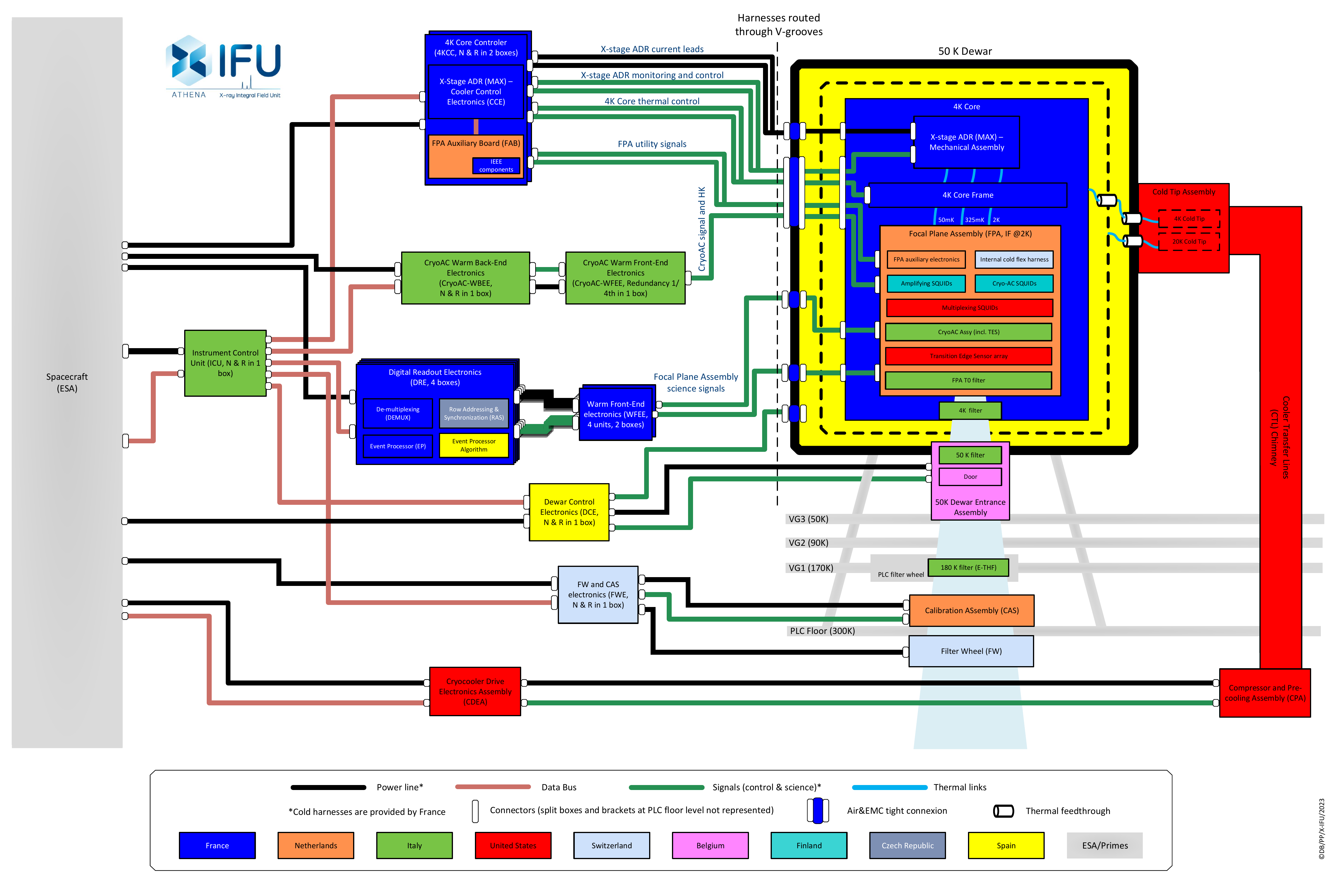}
    \caption{The new physical breakdown of the X-IFU instrument. The main change compared to the previous version is that the 50K Dewar and the 4K cooler are now within the Consortium perimeter and will have to be modeled in the updated LCA. While in the previous incarnation of X-IFU, the Consortium was responsible for the delivery of a 2K core, the new configuration comes with the increased responsibility of having the Dewar and 4K core integrated before delivery.}
    \label{fig:new_phy_breakdown}
\end{figure}

Since the completion of this first analysis, X-IFU entered a redesign phase, leading to changes in the instrument baseline and in the perimeter of activities of the Consortium (see Figure \ref{fig:new_phy_breakdown}). The most notable changes are that the Dewar is now under Consortium responsibility. Similarly, the cooling chain, thanks to the use of passive cooling has been replaced by a single 4K cooler and the last stage by a multi-stage Adiabatic Demagnetization Refrigerator (ADR) cooler (the cooling chain, with the exception of the last stage cooler, and the Dewar were previously in the perimeter of responsibility of ESA, hence excluded from our LCA, explaining why it does not appear in Figure \ref{fig:phy_breakdown}). Most other sub-systems have changed very little compared to the previous instrument design. Yet changes in the performance requirements for X-IFU have led to reducing the number of readout channels, hence the number of electronic boxes, e.g. the Digital Readout Electronics. The current LCA will thus be updated to account for the new perimeter of the Consortium. Some elements, e.g. the 4K cooler, considered today as a procurement by NASA, may be protected by non-disclosure agreements, hence their environmental impacts will have to be approximated. It is hoped that all the other elements making up Athena will also be subject to a LCA, so that the global environmental footprint of the mission is assessed, and further impact reductions, beyond those of X-IFU, identified.

\section{Conclusions}
For the first time for a space astronomy instrument, a LCA was performed to estimate the environmental impact associated with its development. This LCA identified model manufacturing, testing, and business traveling as environmental impact hot spots on which reduction actions should be focused. Among the impacts categories, climate change and resource use, including fossil, mineral and metal resources, turn out to dominate the environmental impacts, a finding that is similar to results obtained for ground-based science facilities \cite{Vargas_2023arXiv230912282V}. More detailed analysis is required to prioritize the actions, e.g. a detailed breakdown of the impacts associated with the use of clean room facilities, across the whole consortium. Nevertheless, in order to verify the efficiency of these actions, a mechanism needs to be put in place that traces their effectiveness from the recommendations to their implementation. 
So far, no such mechanism exists, neither at X-IFU Consortium level, nor at CNES or ESA project level.
One possible solution could be that a dedicated ``environmental architect'', under the responsibility of the CNES X-IFU project manager, overlooks and tracks the implementations of the recommendations.
The architect could also have the responsibility to update the LCA, as the maturity of the instrument design increases and real hardware is produced and tested.
The person may also act as liaison with similar personal in other consortium institutes.

It remains to be seen by what amount environmental impacts can actually be reduced based on LCA-informed action plans.
Studies on eco-design of space missions suggest that environmental impact reductions of several tens of a percent may be achievable, yet that in general, improvements in some environmental indicators can lead to a degradation of others \cite{vercalsteren2018}.
Therefore it seems likely that eco-design will not be sufficient to make space sciences sustainable, and that more significant changes are needed in the modus operandi of the community.
Specifically, LCA need to inform implementation decisions of new space missions, making sure that the limits of planetary boundaries are no longer exceeded.
Of course, space sciences do not exceed these boundaries alone, yet they contribute together with all other human activities that the Earth system no longer resides within the safe operating range \cite{richardson2023}.
We are convinced that vibrant and exciting space sciences are also possible with a more responsible and sober use of resources, by favoring collaboration over competition, stopping the race for ever more and bigger facilities, by enhancing the mining of existing data archives that so far remain often poorly explored, and by investing more time and effort in research and development activities that path the way to new ground-breaking instruments, instead of constantly pushing for new instruments with only incremental improvements, owing to the lack of sufficient technological innovation.

\section{Acknowledgments}
The authors are grateful to the X-IFU sub-system leads that contributed to the data collection used for the LCA, that led to the results presented above. We are also grateful to Bruno Millet (CNES) for supporting the objective to make the development of the instrument more sustainable.  

The authors declare no conflict of interest. The data presented here are available on request. All authors made substantial contributions to the conception or design of the work; the acquisition, analysis, or interpretation of data. They contributed to drafting the work or revised it critically for important intellectual content. They approved the version to be published; and agree to be accountable for all aspects of the work in ensuring that questions related to the accuracy or integrity of any part of the work are appropriately investigated and resolved. This work was funded by CNES, CNRS and CEA in France and by internal resources elsewhere.

We thank the referee for comments on the paper.

\bibliography{db_bibliography.bib}

\end{document}